\documentstyle[11pt,aaspp4]{article}
\begin{document}
\title{A Morphological and Multicolor Survey for Faint QSOs in
the Groth-Westphal Strip\altaffilmark{1,2}}
\author{ Bernhard Beck-Winchatz \altaffilmark{3} and Scott F. Anderson}
\affil{Department of Astronomy, University of Washington, Box 351580, Seattle, WA  98195-1580}
\affil{bbeck@condor.depaul.edu, anderson@astro.washington.edu}

\altaffiltext{1}{Based on observations with the NASA/ESA Hubble Space 
Telescope obtained at the Space Telescope Science Institute, which is 
operated by the Association of Universities for Research in Astronomy, 
Inc., under NASA contract NAS5-26555.}
\altaffiltext{2}{Based on observations obtained with the Apache Point 
Observatory 3.5-meter telescope, which is owned and operated by the 
Astrophysical Research Consortium.}
\altaffiltext{3}{Current address: DePaul University,
Department of Physics, Chicago, Il 60614-3504}

\begin{abstract}

Quasars representative of the populous faint end of the luminosity 
function are frustratingly dim with $m\sim24$ at intermediate redshift; 
moreover groundbased surveys for such faint QSOs suffer substantial 
morphological contamination by compact galaxies having similar 
colors. In order to establish a more reliable ultrafaint QSO sample, we 
used the APO 3.5-m telescope to take deep groundbased $U$-band CCD images 
in fields previously imaged in $V,I$ with WFPC2/HST. Our approach hence 
combines multicolor photometry with the $0.1''$ spatial resolution of HST,
to establish a morphological and multicolor survey for QSOs extending 
about 2 magnitudes fainter than most extant groundbased surveys. We 
present results for the ``Groth-Westphal Strip", in which we identify 
10 high likelihood UV-excess candidates having stellar or 
stellar-nucleus+galaxy morphology in WFPC2. For $m_{606}<24.0$ (roughly 
$B \lesssim 24.5$) the surface density of such QSO candidates is 
$420^{+180}_{-130}$~deg$^{-2}$, or a surface density of 
$290^{+160}_{-110}$~deg$^{-2}$ with an additional $V-I$ cut that may 
further exclude compact emission line galaxies. Even pending
confirming spectroscopy, the observed surface density of QSO candidates 
is already low enough to yield interesting comparisons: our measures 
agree extremely well with the predictions of several recent 
luminosity function models.

\end{abstract}

\newpage
\section{Introduction}

The optical $logN-logS$ curve and ultimately the luminosity function
constrain global models for the evolution of the QSO population (e.g., 
Schmidt \& Green\markcite{Sch83} 1983; Hartwick \& Schade\markcite{HS90} 
1990). The QSO luminosity function has been described via many model 
parameterizations, but is often characterized as relatively flat at the 
populous faint end,  with a steep fall-off to higher luminosities (e.g., 
Boyle, Shanks, \& Peterson\markcite{BSP88} 1988). Physical models 
for the evolution of the QSO population must explain behavior manifest in 
the luminosity function and $logN-logS$ curve. In turn, the observed 
luminosity function and $logN-logS$ curve may constrain parameters in 
physically-based evolution models, such as black hole masses, accretion 
rates, the fraction of galaxies that undergo a QSO phase and the lifetime 
of this phase, luminosity efficiencies and the role of advection-dominated 
flows, etc. (e.g., Blandford\markcite{Blan86} 1986; 
Cavaliere \& Padovani\markcite{Cav89} 1989;  Yi\markcite{Yi96} 1996;
Haiman \& Menou\markcite{Hai99} 1999).

Despite the impressive depth of well known photographic samples for QSOs 
which extend to as faint as $B<22.6$ (e.g., Koo, Kron \& 
Cudworth\markcite{KKC86} 1986), there is a need for surveys for even fainter 
QSOs. Most 
fundamentally, only ultrafaint QSO samples can provide strong constraints 
on the faint, populous end of the luminosity function at intermediate to 
high redshifts. For example, a QSO with $M_B=-22$, i.e., a high space 
density object from the faint portion of the luminosity function, 
will have $B\approx24$ at $z=2.1$ ($H_{0}=50$ km/s/Mpc, $q_{0}=0$). 
In order to probe the faint end of the luminosity function, multicolor 
surveys for ultrafaint QSOs have already been undertaken using groundbased 
CCD searches alone (e.g., see Osmer et al.\markcite{Osm98} 1998 and 
references therein). The multicolor selection approach is similar to 
that used by, for example, Koo et al.\markcite{KKC86} (1986), and
generalized by Warren et al.\markcite{War87} (1987) to identify even 
$z>4$ QSOs. 

While such multicolor  information is excellent for segregating QSOs from 
normal stars, a residual uncertainty in groundbased surveys is the difficulty 
of {\it morphologically} distinguishing very faint stars/QSOs from very 
faint galaxies, in the first place. The potential severity of this problem may 
be realized by noting that there are $\sim 10^{2}$ times as many galaxies
as QSOs per square degree to $m\lesssim 24$; so even a very small 
fraction of faint contaminating galaxies can dominate uncertainties in QSO 
candidate selection. Groundbased surveys for ultrafaint QSOs unavoidably suffer
from such ``morphological" contamination; typical faint field galaxies to 
$V \lesssim 25$ have a median half-light radius of $<1''$ (e.g., Roche et 
al.\markcite{Roc96} 1996), and therefore cannot be readily resolved from the 
ground. Such morphological contamination is likely an obstacle even in current 
state-of-the-art faint groundbased CCD surveys such as the ``Deep Multicolor 
Survey" (hereafter, DMS; e.g., Hall et al.\markcite{Hall96} 1996,
Kennefick et al.\markcite{Ken97} 1997, Osmer et al.\markcite{Osm98} 1998). 
Follow-up spectroscopy 
of the DMS candidates extending to $B\sim22.6$, confirms about 20$\%$ of 
the candidates are in fact QSOs, and the situation at fainter
magnitudes in groundbased surveys is likely to be much worse.
On the other hand, many moderate-redshift galaxies are by comparison
readily morphologically recognizable as resolved objects in 
{\it Hubble Space Telescope} (hereafter, HST) images,
even beyond $m\sim24$ (e.g., Griffiths et al.\markcite{Gri94} 1994).

We describe a program to obtain a combined morphological and multicolor survey 
for faint QSOs and related AGNs to $m \lesssim 24$. Our approach is to
take deep groundbased $u^*$-band (similar to $U$)
CCD frames in fields previously imaged in redder $F606W$ and $F814W$ 
filters from HST. Our survey hence may combine multicolor photometry 
with the $0.1''$ spatial resolution of HST. While the 
multicolor aspect of such a survey is well-known and tested from 
groundbased studies, it is the exquisite morphological information at 
ultrafaint magnitudes that permits HST to make a new contribution.
It should be noted that Sarajedini et al.\markcite{Sar96} (1996, 1999ab) have 
undertaken a morphological survey for low-luminosity AGN searching for 
resolved galaxies with stellar-nuclei in red $F606W$ and $F814W$ WFPC2 images, 
although their data set does not include the $U$-band color information 
traditionally employed to efficiently select QSOs. Moreover,
Conti et al.\markcite{Con99} (1999) and Jarvis \& MacAlpine\markcite{JV99} 
(1999) have recently discussed an approach similar to ours 
that combines color and HST morphology information. Jarvis \& MacAlpine 
concentrate on high-redshift ($z>3.5$) quasar candidates in the Hubble Deep 
Field (HDF), while the Conti et al. study of the HDF also permits an upper 
limit on the surface density of lower redshift QSOs as well, and thus we
discuss the latter results further in \S 5.

Our paper is organized as follows. We discuss the observations and data 
reductions in \S 2, and describe the QSO sample and its selection in \S 3. 
We consider some completeness and contamination issues in \S 4, estimate 
$logN-logS$ curve information in \S 5 and provide 
a comparison of our results with extrapolations from 
brighter existing surveys and predictions based on various luminosity 
function models. Finally, \S 6 
presents a brief summary and concluding remarks. A more detailed account of
much of the information provided herein may be found in 
Beck-Winchatz\markcite{BW98} (1998).

\section{Observations and Reductions}

\subsection{CCD Observations and Basic Reductions}

The most uniform high galactic latitude WFPC2 archival data sets with 
substantial sky coverage employ $F606W$ and $F814W$ filters. Each WFPC2 field 
covers 4.4 arc-min$^2$, is very deep (better than 10\% photometry at 
$m\approx$ 25), and provides superb $\sim$0.1$\arcsec$ angular 
resolution. Such red WFPC2 images are available in the archive for
several hundred fields at high galactic latitude. Most of 
these fields are disjoint and widely spread across the sky, but
the ``Groth-Westphal Strip'' (Groth et al.\markcite{Gro94} 1994) 
conveniently and efficiently images a reasonably large adjacent area of sky; 
this adjacency substantially enhances some aspects of complementary 
groundbased observations (e.g., see also Koo et 
al.\markcite{Koo96} 1996; Brunner, Connolly, \& Szalay\markcite{Bru99} 1999).

HST WFPC2 observations of the Groth-Westphal Strip were carried out 
during March and April of 1994 (Groth et al.\markcite{Gro94} 1994), and 
include a mosaic of 
28 slightly overlapping adjacent WFPC2 fields imaged in both $F606W$ (with 
stacked exposure times of about 2800~s) and $F814W$ (with stacked exposure 
times of about 4400~s). Bias subtraction, flatfield correction, etc., were 
performed by the standard STScI reduction pipeline. Cosmic ray removal and 
coaddition of the 4 separate exposures in each WFPC2 field and bandpass were 
carried out by the authors. No correction for the charge-transfer efficiency 
(CTE) problem was applied, since the error introduced  is expected to be small 
(e.g., Holtzman et al.\markcite{Hol95} 1995). Vignetted regions of the WFPC2 
images were trimmed before analysis.

The $F606W$ and $F814W$ filters alone do not provide adequate color information
to efficiently select quasar candidates, so we obtained complementary
bluer CCD images of the Groth-Westphal Strip, remotely using the 3.5-m 
Apache Point Observatory (APO) telescope (owned and operated by the
Astrophysical Research Consortium) in New Mexico. 
The APO 3.5-m data discussed here were taken with the Seaver Prototype Imaging 
CCD camera (hereafter SPIcam), using a filter which we refer to here
as $u^*$; this filter is similar to the Sloan Digital Sky Survey (SDSS)
$u'$ filter (e.g., Fukugita et al.\markcite{Fuk96} 1996). SPIcam is a backside 
illuminated SITe 2048$^2$ pixel 
device with 24~$\mu$ pixels, and a plate scale of 0.28$''$ per pixel with 
2$\times$2 binning. Bias subtraction, flatfield correction,
trimming, and cosmic ray removal were performed in standard fashion. 
The SPIcam field of view can capture most (about 90\%) of 3 adjacent WFPC2 
fields in a single image, and our survey region concentrates on 
24 adjacent WFPC2 fields of the Groth-Westphal Strip, which 
we have also imaged in 8 $u^*$ SPIcam fields. (Our survey region does not
include the four most NE WFPC2 fields of the Strip). Typically, each of the 
8 SPIcam fields was observed in $u^*$ with a total exposure time of about 
8000~s.

Our survey area has a complicated geometry of partially overlapping 
WFPC2 and SPIcam fields. We calculate the survey region to be 0.0245 deg$^2$ by 
projecting the WFPC2 fields onto the SPIcam fields, 
and adding up all SPIcam pixels which overlap an unvignetted portion of at
least one WFC field (accounting for areas covered more than once).

\subsection{Object Detection and Photometry in Stacked CCD Frames}

Object detection in the stacked WFPC2 images was performed with 
SExtractor V1.2 (Bertin \& Arnouts\markcite{BA96} 1996), generating a 
catalog of about 6700 objects; detection completeness is very high to our 
survey magnitude limits (see \S 4.1). The  Groth-Westphal Strip is at high 
Galactic latitude, 
and since crowding is not a problem, aperture photometry was obtained for all 
objects using a 3 pixel radius aperture. Aperture corrections were performed 
according to the prescription by Holtzman et al.\markcite{Hol95} (1995). 
Zeropoints for the Vega magnitude system were taken from 
Whitmore\markcite{Whit95} (1995). The HST Medium Deep Survey project 
(hereafter MDS; Ratnatunga, Griffiths, \& Ostrander\markcite{Rat99} 1999) has 
electronically published a catalog of objects in the Groth-Westphal Strip 
which includes ``total maximum likelihood estimation'' $F606W$ and 
$F814W$ magnitudes for each object. The MDS catalog thus provides an 
independent check on the WFPC2 detection, photometry, and morphological 
classification algorithms we employ (see below).
For stellar-PSF objects within our survey limits of $19.3<m_{606}<24.0$, we 
find excellent agreement between our photometry and that of the MDS:
$m_{606}^{ap} - m_{606}^{mds}=0.09\pm0.11$, and 
$(m_{606}-m_{814})^{ap} - (m_{606}-m_{814})^{mds}=0.03\pm 0.10$, where the 
superscripts $ap$ and $mds$ denote our aperture photometry and MDS photometry, 
respectively. 

Object detection in the groundbased $u^*$ SPIcam images was also performed with 
SExtractor, generating a catalog of about 1700 objects (see \S 4.1 for
quantification of detection completeness). Instrumental magnitudes for all
SPIcam detected objects were calculated in a 2.8$\arcsec$ diameter aperture;
aperture corrections were estimated
using the 4-5 brightest unsaturated stellar-PSF objects in each stacked image
and should be accurate to $<$0.05 magnitudes. Calibration to a
$u^*$ system similar to the expectations for the eventual $u'$ SDSS 
system (the SDSS system is currently being carefully defined) were carried 
out via observations of Feige 34, Feige 66, BD +26$^{\rm o}$ 2606, and 
BD +33$^{\rm o}$ 2642, and a number of Landoldt standards (using 
approximate transformations in Fukugita et al.\markcite{Fuk96} 1996), 
and adopting $u^*$=10.78 for BD+26$^{\rm o}$ 2606.

Astrometric information is available for the WFCP2 images via the STScI 
pipeline reductions; a small correction was made to the astrometric 
information in the original headers using the {\it STSDAS} script called 
`uchcoord' (version 1.1). An astrometric solution was also derived for each 
stacked 
SPIcam image using the Digitized Sky Survey. Objects detected in both WFCP2 
and SPIcam images, whose positions agreed (before corrections) to better than 
2$\arcsec$, were used to place the two data sets on the same astrometric 
frame. After correction, the mean deviation between positions
of stellar-PSF objects derived from WFPC2 versus SPIcam frames is 0.4$\arcsec$.

\section{Sample Selection}

The terminology ``quasi-stellar" of course traditionally has implied 
approximately 
stellar morphology in groundbased images. The morphological discrimination in 
groundbased images of bright stars and QSO candidates on the one hand from 
bright galaxies on the other is relatively straight-forward. However, faint 
field galaxies typically have half-light radii $<1\arcsec$ 
(e.g., Roche et al.\markcite{Roc96} 1996) 
and cannot readily be resolved from the ground.  Our survey aims to reduce the 
morphological contamination inherent in deep groundbased multicolor surveys by 
taking advantage of the 0.1$\arcsec$ resolution of HST/WFPC2 images in the 
survey region. Broadly speaking, our method is to select and then merge two 
subsamples of UV-excess QSO candidates.
We first select a sub-sample of QSO candidates 
consisting of UV-excess objects with $19.3<m_{606}<24.0$ that are unresolved 
(i.e., have stellar-PSF morphology) in the WFPC2 images.
Then as the host galaxies of QSOs may be resolved at the angular 
resolution of HST (e.g., Hutchings et al.\markcite{Hut94} 1994;
Bahcall et al.\markcite{Bah97} 1997; Hooper et al.\markcite{Hoo97} 1997), we 
also separately select UV-excess objects that appear to have stellar-PSFs in 
our {\it groundbased} SPIcam images, and then further examine their morphology 
in detail in the WFPC2 images; we ultimately retain in the second subsample 
only those objects whose morphology based on quantitative criteria is 
consistent with an approximately unresolved (stellar-PSF) nucleus in WFPC2, 
even if an underlying galaxy is also evident. Our final QSO sample 
consistently merges candidates from these two UV-excess sub-samples.

\subsection{QSO Candidates with Stellar-PSF Morphology in WFPC2}

We first select objects having stellar morphology in WFPC2 via application
of the SExtractor software to the 72 stacked WFC images of our
survey region of the Groth-Westphal Strip. SExtractor, which uses a neural 
network for stellar-PSF vs. galaxy separation, outputs a ``stellarity index"
for each object. This index is 0.0 and 1.0 for objects that are 
confidently identified as galaxies and stars, respectively; more ambiguous 
objects are assigned values close to 0.5. An input parameter to 
SExtractor that influences this output stellarity index is a user-specified 
FWHM of stellar-PSF objects.
We estimate an appropriate input FWHM iteratively. We start by  fitting a 
Gaussian to a few high signal-to-noise stellar (in WFPC2) objects to determine 
a representative FWHM. All 72 WFC images are then analyzed with SExtractor 
using this initial input FWHM value. We then fit a 
Gaussian to all objects with a stellarity index greater than 0.5 and  use the 
median FWHM of Gaussian fits from that iteration to reanalyze the
images once again with SExtractor, etc. This process is repeated
iteratively and converges to finally adopted SExtractor input FWHM values of 
1.48 pixels for $F606W$ images and 1.60 pixels for $F814W$ images. We verify 
that these are reasonable values first by visually confirming that
high signal-to-noise stars and galaxies, and low signal-to-noise ambiguous
objects are finally assigned stellarity indices close to 1.0, 0.0, and 
0.5, respectively. Objects with SExtractor stellarity indices greater than 
0.5 in both WFPC2 filters are ultimately considered morphologically
``stellar" in our sample. Second, we appeal to an entirely independent 
morphology algorithm as described below as a further quantitative check that we 
have not missed truly stellar-PSF objects.

Shown in Figure 1 is a two-color diagram of objects 
with $19.3<m_{606}<24.0$ found to have stellar morphology in 
$F606W$ and $F814W$ WFPC2 images according to the SExtractor analysis. 
For added clarity in Figure 1, we have not displayed data for 
non-detections in $u^*$ (with one exception discussed below), although
objects with such limits are accounted for in our analyses. It is 
reassuring that the bulk of the objects in this two-color diagram appear to 
indeed comprise a ``stellar sequence", empirically confirming that the 
morphological criteria employed have excluded most galaxies. A handful of 
odd-color outliers, including quasar candidates (solid symbols) having 
stellar-morphology in WFPC2, and perhaps a few still-unresolved compact 
galaxies, are also evident in Figure 1; we discuss these below.

Figure 2 duplicates the data in Figure 1, except that we also overplot
for approximate comparison
the synthetic colors expected for main sequence stars M8 through F8
(5-point star symbols in Figure 2). To generate these synthetic
points, we use the Gunn \& Stryker\markcite{GS83} (1983) catalog 
and {\it SYNPHOT} to convolve catalog spectrophotometry with 
the wavelength dependent sensitivity curves of WFPC2 $F606W$/$F814W$ 
and an approximate $u^*$ sensitivity curve.
There is a reasonably good match between observed and synthetic 
stellar sequences, especially as the $u^*$ sensitivity curve is only an 
approximate prediction; for example, the available $u^*$ transmission curve is 
actually for a different detector and telescope, the low metallicity expected 
for halo stars has not been accounted for, etc. There appear to be few main 
sequence stars earlier than early-G/late-F in our ``stellar" sample as 
might be predictable for faint halo stars; indeed a clump of likely
low-metallicity Galactic halo subdwarfs appears as expected near 
$(m_{606}-m_{814})\sim0.55$, but more ultraviolet than the main sequence. 
For additional rough comparison, we also show in Figure 2 the colors
expected for a ``typical" QSO at various redshifts (solid curve). 
We estimate these colors using the composite 
spectrum derived from 700 QSOs of the Large Bright Quasar Survey 
(LBQS; Francis et al.\markcite{Fran91} 1991), redshifted and convolved 
with the approximate $u^*$, $F606W$, and $F814W$  sensitivity curves. 
As may be verified in this figure, typical QSOs with redshifts $z<$2.1
appear more ultraviolet in ($u^*-m_{606}$) than normal stars. 

Traditional UV-excess surveys have selected objects with $(U-B)<-0.4$, a 
criterion often thought to recover about 95\% of $z<2.2$ QSOs 
(Veron\markcite{Ver83} 1983); at larger redshifts 
Ly-$\alpha$ leaves the $U$-band and enters the $B$-band, with $(U-B)$ 
becoming redder rapidly and assuming
the values of normal stars.  The red edge of the $u^*$-filter we use is at a 
slightly shorter wavelength than the red edge of the $U$-band, and consequently 
our redshift limit is slightly lower at about $z\lesssim 2.1$ We adopt a 
UV-excess limit of $(u^*-m_{606})<0.7$ for QSO candidate selection; for 
reference this limit is shown by the horizontal dotted line in Figures 1--2. 
Note that near the $m_{606}<24$ survey limit there are red objects not detected
in $u^*$, but none of these non-detections have limits in  
$(u^*-m_{606})$ which satisfy our UV-excess criterion; 
in part, a desire to avoid such complications motivates the choice of our
survey limit of $m_{606}<24.0$. 

The four QSO candidates that meet our criteria (UV-excess and unresolved 
stellar morphology in WFPC2 according to SExtractor) are marked by solid 
circles in Figures 1--2, and are listed in the first four rows of Table 1. 
Upon visual examination of the WFPC2 images, the candidate we catalog as
\#98 does appear to have an associated underlying galaxy. However, the 
dominant nucleus certainly appears stellar;
moreover, this object also satisfies a third set of 
quantitative morphology criteria discussed in \S 3.2., and therefore is 
appropriately retained in the final sample in any case.

As noted above, the Medium Deep Survey project (Ratnatunga et 
al.\markcite{Rat99} 1999) has 
developed  two-dimensional maximum likelihood image analysis software and
previously applied it to the WFPC2 images of the Groth-Westphal Strip. 
Because their algorithms for detection and morphological star-galaxy 
separation are entirely independent of the above SExtractor neural network 
approach, the electronic MDS catalogs provide a valuable cross-check on our 
analyses. Objects in the MDS algorithms are morphologically classified as
stellar, disk, bulge, or disk+bulge (MDS classes 0, 1, 2, and 3). 
All 165 objects in our survey area, and within our survey magnitude limits,
which are classified in the MDS catalogs as
morphologically stellar in both the $F606W$ and $F814W$, are
also classified as morphologically stellar by our SExtractor analysis.
Our SExtractor analysis classifies a few more (about 5\%) objects 
as ``stellar" in WFPC2 than the MDS catalogs, so our SExtractor classifications
in WFPC2 appear robust but slightly more inclusive than the MDS ones.
The excellent agreement between SExtractor stellar-PSF classifications and 
those of the MDS confirms the efficacy of both approaches. All UV-excess QSO 
candidates with stellar morphology in WFPC2, were already identified via the 
SExtractor neural network approach; that is, no additional QSO candidates with 
stellar morphology and having UV-excess are added from consideration of 
objects independently classified as ``stellar" in the MDS catalogs. 

We also call attention to two non-UV-excess objects (solid squares in 
Figures 1--2) with stellar-PSF morphology in WFPC2, whose unusual colors 
fall markedly above the stellar sequence and close to the colors predicted 
for $z\sim 3$ QSOs. We list these miscellaneous higher-redshift QSO 
candidates in the last two rows of Table 1, but of course do not include 
them in our UV-excess sample.

\subsection{QSO/AGN Candidates with Stellar-Nuclei in WFPC2, 
but having Resolved Underlying Galaxies}

As noted above, HST studies show that some QSOs with stellar morphology in 
groundbased images have resolved host galaxies when observed at WFPC2 
spatial resolution. QSOs with such host galaxies would not necessarily have 
been classified as morphologically stellar by either the SExtractor neural 
network or the MDS algorithms applied to WFPC2 images as described in \S 3.1. 
Our strategy to find such QSOs with underlying host galaxies is to initially 
identify UV-excess objects that are approximately stellar in {\it groundbased} 
images, and then consider their morphology in more detail using the HST images. 
Ultimately, we will select a subsample of these UV-excess objects for which 
the nucleus is approximately stellar, even if an 
underlying galaxy is also evident in WFPC2.

For the initial stages of the selection of this second subsample, we again 
use morphology information from SExtractor, but this time applied to the 
stacked $u^*$ SPIcam groundbased images. Again, the quality of the 
star-galaxy morphology separation depends on the value input to SExtractor for
the FWHM of stellar objects.  As seeing in the groundbased images differs 
from field to field, we estimate input FWHM values for each stacked SPIcam
image separately, using the median FWHM of the $\sim$10 brightest stellar 
objects in each stacked image. The adopted input FWHM's vary between 5.0 and 
6.6 pixels (1.4$\arcsec$-1.9$\arcsec$), and we verify that the neural
network assigns reasonable output stellarity indices by confirming that high
signal-to-noise galaxies and stars, and low signal-to-noise ambiguous objects,
are assigned values close to 0.0, 1.0, and 0.5, respectively.

The positions of UV-excess objects detected in the $u^*$ groundbased images are 
matched to our catalogs of objects detected in the redder WFPC2 
filters. In our survey area, we find 77 objects that have UV-excess,
$19.3<m_{606}<24.0$, and SExtractor stellarity indices $>$0.5
in the stacked {\it groundbased} $u^*$ images. Although these UV-excess objects
are stellar in the groundbased images (i.e., a traditional definition of
``quasi-stellar"), an initial visual examination of the 77 UV-excess 
candidates shows 
that most (80\%) are readily resolved into galaxies on the WFPC2 images, 
confirming the likely severity of morphological contamination of ultrafaint 
QSO surveys that rely entirely on groundbased imaging data. Because many of
the objects selected in this fashion turn out to be extended in WFPC2, we 
initially consider their MDS ``total magnitudes" in $F606W$ and $F814W$. Such 
model magnitudes may be preferred to WFPC2 aperture magnitudes for 
some extended objects, although when we later confine our attention to just 
that subset of the objects with stellar-nuclei in WFPC2, we will return to 
small-aperture magnitudes for the nuclei.

In order to establish an objectively-selected sub-sample, we employ 
quantitative morphological tests to select QSO/AGN candidates having a compact
approximately-stellar nucleus, but resolved host galaxy in WFPC2.
These morphological tests consist of two parts. 
The first part is designed to exclude 
UV-excess galaxies whose nuclei are significantly more extended 
than typical stellar-PSF objects in WFPC2. We calculate a normalized integrated 
and azimuthally-averaged radial profile parameter $I_{RAD}$ for the core of 
each of the 77 initial UV-excess objects.  We measure the $F606W$-fluxes 
through apertures of radii of one, two, and three pixels and divide these by 
the total flux of the object measured at an aperture radius of 14 pixels.  
For comparison, a template radial profile typical of truly stellar objects is 
derived by averaging the similarly measured and normalized fluxes of the high 
signal-to-noise ($m_{606}<21$) stellar 
objects (as classified by SExtractor) in the WFPC2 images. We then 
integrate the area between this template profile for stellar-PSF's and each
object's core profile (this is the parameter $I_{RAD}$), and exclude all
objects for which the integral is larger than a certain threshold value.
We select a threshold value based on the observed distribution (Figure 3) 
of the integral values $I_{RAD}$ for stellar-PSF objects (as classified by the 
neural network in WFPC2) to the magnitude limit of the survey. With our 
adopted value of $I_{RAD}<$0.5, all but one of the objects 
classified by SExtractor as having stellar-PSF morphology
in WFPC2 would have passed our $I_{RAD}$ test.

A second morphological test is needed however, as some very compact UV-excess 
objects that are slightly elongated in WFPC2 images pass the first 
$I_{RAD}$ radial profile test described above; in part, this is due to the 
azimuthal averaging with $I_{RAD}$. Yet UV-excess objects with such compact 
elliptical morphology are most likely compact narrow emission line galaxies 
(or CNELG's), which are known to be the major contaminant
of groundbased surveys for faint QSOs (e.g., Koo \& Kron\markcite{KK88}
1988; Hall et al.\markcite{Hall96} 1996). We therefore also then test the 
roundness of UV-excess objects
(which pass the $I_{RAD}$ test) in WFPC2 by fitting two-dimensional Gaussians
to their profiles. Objects for which the major to minor axis ratio 
$R_{MAJ}/R_{MIN}$ of the best-fit Gaussian falls above a certain threshold are 
also morphologically excluded from the final QSO candidate sample. We base the 
value of this threshold on the observed distribution (Figure 4) of major to 
minor axis ratios of the stellar-PSF objects selected by SExtractor. With our 
adopted value of R$_{\rm MAJ}$/R$_{\rm MIN}<$1.5, every object classified by
SExtractor as having stellar-morphology in WFPC2 would have passed this
second ellipticity test.

Table 2 lists the seven UV-excess objects initially classified by SExtractor 
as having stellar morphology in our {\it groundbased} images, which also 
pass both our $I_{RAD}$ and $R_{MAJ}/R_{MIN}$ morphology tests in WFPC2 
frames. We list in Table 2 only the additional QSO/AGN candidates found in 
this fashion, and do not repeat those from Table 1 (which have stellar 
morphology in WFPC2, and so also satisfy the $I_{RAD}$ and 
$R_{MAJ}/R_{MIN}$ criteria). 
The objects in Table 2 are QSO/AGN candidates with approximately 
stellar-nuclei, even if they have resolved host galaxies in WFPC2. Their 
location in the two color diagram is shown in Figure 5 (see filled circles);
note that other data in Figures 1 and 5 are quite similar, except that
Figure 5 displays MDS catalog $F606W$ and $F814W$ magnitudes.

Table 3 lists other miscellaneous UV-excess objects that fall near to, but
somewhat outside, the parameter space of our QSO/AGN selection criteria.
This miscellaneous list includes mainly UV-excess objects we might have 
identified as possible QSO candidates via a simple visual examination of the 
WFPC2 images but which fail the quantitative morphology tests,
plus one UV object (\#1176) that meets the morphology criteria but
is slightly fainter than the magnitude limit of the survey.
We do not consider these miscellaneous objects further in the current paper.

\subsection{The Combined Samples of QSO/AGN Candidates}

Our combined sample of QSO candidates---called Sample~I---is conceptually 
the merging of UV-excess objects selected in \S 3.1 and 3.2 and listed in 
Tables 1 and 2, 
and includes objects with stellar morphology in WFPC2, as well as 
objects with approximately-stellar nuclei in WFPC2 even if the 
underlying host galaxies are resolved. However, for greater consistency with 
the stellar-PSF objects selected in \S3.1, in combining the subsamples 
from \S3.1 and \S3.2 we henceforth consider only the small-aperture 
WFPC2 photometry. As quantified above, there is excellent agreement 
between our aperture measures and the MDS ``total magnitudes" for point 
sources, but the total MDS magnitudes for the stellar-nucleus + galaxy 
QSO candidates discussed in \S3.2 tend to be slightly brighter than our 
WFPC2 aperture magnitudes. Even so, this small systematic difference in
photometry for extended objects effectively eliminates only the QSO
candidate \#72 in Table 2 from further consideration in the merged Sample I:
its small-aperture magnitude slightly exceeds $m_{606}=24.0$, although 
the MDS total magnitude is 23.9.

The QSO candidates in Sample I are chosen according to the following 
criteria: aperture magnitudes $19.3<m_{606}<24.0$, $(u^*-m_{606})<0.7$, and 
either (a) SExtractor stellarity indices $>$0.5 in $F606W$ and $F814W$ images, 
or (b) SExtractor stellarity index $>$0.5 in $u^*$ groundbased images, and 
meeting integral $I_{RAD}$ and roundness $R_{MAJ}/R_{MIN}$ criteria for 
stellar-nuclei objects in WFPC2. Table 4 provides merged summary information
for the 10 QSO candidates of Sample~I.

As discussed earlier, compact narrow emission line galaxies (CNELG's) are 
known to have $(U-B)$ colors similar to those of $z\lesssim 2$ QSOs but often 
with somewhat redder $(B-V)$ indices. They are a major contaminant of 
groundbased surveys for faint QSOs. Some other multicolor surveys thus have 
attempted to reduce this contamination by excluding UV-excess objects redder 
than a certain $(B-V)$ from their candidate list. For example, 
Hall et al.\markcite{Hall96} (1996) apply a magnitude dependent cut-off 
value $(B-V)$=0.25-0.4, while Koo et al.\markcite{KKC86} (1986) exclude 
objects with $(J-F)>$0.55. We have already 
attempted to exclude CNELG's on the basis of their morphology, but to
reduce residual contamination we also devise a second sample---which we 
call Sample II---with a red limit on the $(m_{606}-m_{814})$ color, in 
addition to the UV-excess and morphology criteria described above. 
Using synthetic photometry applied to the LBQS composite QSO spectrum 
redshifted through the range $z$=0.5-2.1, we estimate 
$\langle m_{606}-m_{814} \rangle \approx \langle B-V \rangle $+0.2.
We thus adopt a red limit of $(m_{606}-m_{814W})<$0.55 for Sample II, 
corresponding to $(B-V) \lesssim 0.35$ for typical QSOs. Sample II includes 
seven objects: three QSO candidates, \#1148, 507, and 245, from the full 
Sample I are redder than this $(m_{606}-m_{814})$ limit,
and are excluded from Sample II.

\section{Some Completeness and Contamination Issues}

As we will show \S 5, the surface density of {\it candidates}
we find is extremely well matched to expectations for QSOs based on
extrapolations from surveys at brighter magnitudes, and predictions of
various model luminosity functions; this suggests the possibility of high
completeness and low contamination in our sample, although definitive
confirmation must await follow-on spectroscopy. The completeness of UV-excess 
selection is addressed by many other studies (e.g., Veron\markcite{Ver83} 
1983) and as noted above is generally thought to be 90--95\% for 
$z\lesssim 2.2$ QSOs (but see Graham, Clowes, 
\& Campusano\markcite{Gra98} 1998 for an alternate view). Here we 
address a few other issues bearing on completeness/contamination. 

\subsection{Detection Completeness}

We use Monte Carlo simulations to establish the completeness of the SExtractor 
detection algorithm as a function of $u^*$ magnitude in our groundbased
SPIcam images. This is relevant for QSO/AGN candidates selected in \S 3.2.
Artificial ``stellar" (in 1.6$''$ groundbased images)
objects of known magnitudes are added to these SPIcam images, which are then 
reanalyzed using SExtractor. The PSF of the artificial objects is modeled 
using the brightest actual stellar objects in each field. We detect $\sim$80
``stellar" objects to $u^*$=25 in a typical SPIcam field, and find that adding 
8 artificial objects per field does not significantly alter SExtractor
detection or classification of the real objects. A total of 288
artificial objects per 0.$^{\rm m}$1 bin were created between $u^*$=23.0 and  
$u^*$=25.0 and added to each of the stacked SPIcam images.
Positions of artificial objects on the stacked SPIcam frames 
were assigned by a random number generator. Poisson noise was added
to the artificial star template PSF, while readnoise and background noise are 
already accounted for by adding the artificial objects to the existing 
CCD images. The artificial images were then analyzed with SExtractor
in the same manner as the real data.
All 8 of our stacked SPIcam images reach similar depths.  We show in 
Table 5 the detection probabilities based on these tests.
(We did not perform extensive Monte Carlo simulations for 20$<u^*<$23,
but expect detection probabilities $>0.97$ by extrapolation from
our simulations at $u^*>$23). Our survey limit of $m_{606}<24.0$ is also chosen
to insure that SPIcam $u^*$ detection completeness typically exceeds about 
$50\%$ even for the least UV-excess objects that might potentially enter the 
sample at the survey limits (and for the bulk of the sample objects, 
detection completeness in $u^*$ images is in the 80-90\% regime).

\newpage

Both our own completeness tests for our SExtractor derived WFPC2 catalogs, and 
tests independently carried out by the creators of the MDS catalogs 
(Ratnatunga et al.\markcite{Rat99} 1999), confirm that object detection in 
the even deeper WFPC2 images is nearly 100\% complete to at least 
a half-magnitude fainter than our survey limit.
Thus, no completeness corrections are made for those UV-excess QSO candidates 
(discussed in \S 3.1 and listed in Table 1) initially selected from the catalogs
of WFPC2 objects having stellar morphology.

\subsection{Contamination by Stars}

While the major contaminant of UV-excess quasar surveys at bright 
limiting magnitudes are galactic stars (e.g., hot white dwarfs and 
low-metallicity subdwarfs), deeper UV-excess and multicolor QSO surveys 
with confirming spectroscopy show that to $B\sim$22.5 stars are generally 
minor contaminants. For example, in the multicolor selected 
sample of QSO candidates in SA57 (Koo et al.\markcite{KKC86} 1986;
Koo \& Kron\markcite{KK88} 1988), 10\% and 15\% of the  candidates at 
$B\le$22 are respectively white dwarfs and subdwarfs, while at $B>$22 these 
stars make up only 4\% and 7\%
of the candidates, respectively. At most, only 3 of 10 of our UV-excess 
candidates might be stars, and indeed one of these three objects, \#245,
may also weakly show evidence for a host galaxy in $F814W$; 
the other 7 candidates in Sample I are morphologically resolved according to 
the algorithms and/or in a visual examination of the WFPC2 images.

It is unlikely that all three potentially stellar in WFPC2 QSO candidates 
(\#971, 815, and 245) are white dwarfs. The absolute magnitudes of disk white 
dwarfs may be estimated from their colors (Fleming et al.\markcite{Fle86} 1986) 
as $M_V \sim {\rm 11.9} + {\rm 2.9}(U-V) - {\rm 0.5}(U-V)^{\rm 2}$.
We estimate $(U-V)\sim -0.4$ for \#971 and $\sim-0.7$ for 
\#815 and \#245, using synthetic photometry to approximately convert 
between our magnitudes and the Johnson system. Thus candidates
\# 971, 815, and 245 would have $M_V\approx$10.7, 9.6, 
and 9.6, respectively, were they white dwarfs. Assuming a 
disk scale height for white dwarfs of $\lesssim$500 pc, one might expect 
$d\lesssim$1000~pc for disk white dwarfs toward the Groth-Westphal Strip 
($b$=60$^{\rm 0}$). However, the apparent magnitude of a white dwarf with 
$M_V$=9.6 seen at a distance of 1000 pc is $V\approx 19.6$, and therefore 
candidates \#815 and \#245 are much too faint to be disk white dwarfs. 
The object cataloged as ID \#971 would be at a more plausible (although still 
distant) $\sim$600~pc above the galactic plane; and it is not inconceivable 
to expect $\lesssim1$ disk white dwarf in our survey area. 
Assuming $d\lesssim$1000 pc and for our survey area of 0.025 deg$^{\rm 2}$, we
sample a volume of the disk for white dwarfs of $\sim$2500 pc$^{\rm 3}$. The 
space density of hot white dwarfs to $M_V<$11.25 is 
$\sim1.5 \times$10$^{\rm -4}$ pc$^{\rm -3}$ 
(Liebert, Dahn, \& Monet\markcite{Lie88} 1988), and hence, 
the total number of hot (UV-excess) disk white dwarfs expected in our survey 
is $\lesssim0.4$. (see also Gould, Flynn, \& Bahcall\markcite{Gou98} 
1998).

One might also entertain some possibility that \#815 and \#245
are very distant white dwarfs. Objects having the 
$(u^* - m_{606})$ colors of \#815 and \#245, were they similar to disk white 
dwarfs, would be at $\sim$8 kpc
and well into the halo. However, on rather general grounds, it might be 
surprising to find a significant population of very blue white dwarfs abundant 
in the halo population (although see Beck-Winchatz\markcite{BW98} 1998 for 
further discussion, and e.g., Hansen\markcite{Han98} 1998).

Contamination by subdwarfs is difficult to dismiss. The subdwarf clump near the 
location of late-F/early-G stars with bluer $(u^*-m_{606})$ colors 
than main sequence stars appears to be evident in our two-color diagrams (e.g., 
Figure 1). All stellar-PSF QSO candidates in our sample are more UV-excess 
than this clump. It seems unlikely these QSO candidates are actually subdwarfs, 
but follow-on spectroscopy is certainly desirable to confidently confirm their 
nature.

\subsection {Morphological Contamination and Incompleteness}

Central to our approach is the use of HST morphology information to 
select a subset of the UV-excess objects that have either stellar-PSFs
or which have resolved galaxies with approximately stellar-nuclei; this
morphological selection should (and appears to) eliminate  
a large fraction of the compact galaxies that contaminate 
groundbased surveys for faint QSOs. However, there may of course still be 
some residual contamination by extremely compact galaxies inadequately
resolved even at HST resolution. We have attempted to further reduce such 
contamination by generating Sample II which excludes three Sample~I objects on 
the basis of their red $(m_{606}-m_{814})$ color. However, a full assessment 
of the contamination of our sample by non-active galaxies awaits spectroscopic 
follow-up. 

As noted above, among the objects with apparently stellar PSFs in our 
{\it groundbased} images there are 77 which have a UV-excess.
One might also consider if within this group there are a significant number of 
QSOs ultimately excluded by our WFPC2 morphology criteria of \S 3.2. One 
indication that this is unlikely to be a major (e.g., more than 2$\times$) 
effect is provided in the recent HST study by 
Malkan, Gorjian, \& Tam\markcite{Mal98} (1998) of the morphology of nearby 
Seyferts as revealed by WFPC2 images. Among nearby Seyfert 1 galaxies,
Malkan et al. find that only 37\% show no discernible point sources, or have 
point sources which contribute less than $\sim$45\% of the light within the 
inner one arc second. Nonetheless, the Malkan et al. study leaves
open the possibility that some 
of the UV-excess objects lacking PSF-nuclei in WFPC2 (which we have 
excluded from our samples) are in fact QSOs or Seyferts~1s. Thus,
spectroscopic follow-up of some of these morphologically excluded objects 
would also be useful.

\section{Surface Density of Ultrafaint QSO/AGN Candidates}

While the luminosity function provides a more direct test,
many checks of statistical and physical models of QSO evolution can be 
made using observed number-magnitude relations alone. Here we derive 
the $logN-logS$ curve for QSO {\it candidates} to $m<24$. Pending 
confirming spectroscopy, of course, we 
cannot be sure that our candidates are indeed QSOs/AGNs. However as
we now show, our derived QSO candidate surface density at ultrafaint
magnitudes is sufficiently low to 
already provide interesting comparisons with extrapolations from the 
$logN-logS$ curve of brighter quasars, as well as various
model luminosity function predictions.

While our survey is $F606W$ magnitude limited, most existing QSO surveys
define magnitude limits and quote surface densities in terms of Johnson $B$
magnitudes. In order to roughly compare our results with those 
other surveys, we thus first derive an approximate transformation from 
$F606W$ magnitudes to the $B$-band (Figure 6). 
We model a typical QSO spectral shape by fitting a cubic spline to the 
continuum windows identified by Francis et al.\markcite{Fran91} (1991) at 
1285, 2200, 4200, and 
5770~\AA\ for the LBQS composite spectrum. Then both the LBQS composite 
spectrum itself and the continuum fit are numerically redshifted and convolved 
with the $F606W$ and $B$-bandpasses, to yield a rough conversion between 
observed (with emission lines) $F606W$ magnitudes and continuum (without 
lines) $B$-band magnitudes. This conversion, $(m_{606}-B)$, as a function of 
redshift is shown in Figure~6. 
Since the $F606W$ and $B$ passbands cover markedly different wavelength ranges, 
the derived transformation is highly dependent on the assumed spectrum and 
should not be used for any objects except for $z\lesssim$2.1 QSOs, and even then
only for rough comparison.  
Since we do not know the redshifts of the QSOs among our candidates, we adopt 
a typical transformation of $B\approx m_{606}+0.5$. 

Table 6 shows the cumulative surface densities implied by our QSO survey at 
ultrafaint magnitudes. The third and fifth columns list the number of 
candidates
in Samples~I and II, corrected for detection completeness as per \S4.1.
Root $n$ error bars can be highly misleading for such small numbers, and so
errors are estimated from the Poisson fiducial limits appropriate for small 
number counts tabulated by Regener\markcite{Reg51} (1951). Columns four and 
six give the surface densities, corrected for detection incompleteness. 
The cumulative surface density of QSO candidates for 
$19.3< m_{606} <24.0$---the approximate equivalent of 
$19.8 \lesssim B \lesssim 24.5$---is 420$^{\rm +180}_{\rm -130}$ deg$^{-2}$ for 
Sample~I,  or 290$^{\rm +160}_{\rm -110}$ deg$^{-2}$ for Sample II.

In Figure 7 we plot these cumulative surface densities for our
ultrafaint candidates (large filled squares)
along with comparison points from other brighter QSO surveys and several 
predictions of model luminosity functions for $z<2.1$.
The $logN-logS$ curve data derived from other brighter low-$z$ QSO surveys 
(mainly UV-excess) are taken from Hartwick \& Schade\markcite{HS90} (1990; 
hereafter, HS) who compiled data for $\sim10^3$ confirmed QSOs in the 
magnitude range 12.5$<B<$22.5. As discussed above, the 
somewhat bluer edge of the $u^*$ passband as compared to the Johnson $U$ 
passband limits the sensitivity of our survey to $z\lesssim$2.1. We estimate 
$N$($z<$2.1)/$N$($z<$2.2)$\approx$0.96 by interpolating 
the redshift distribution of multicolor selected QSOs derived by 
Koo \& Kron\markcite{KK88} (1988), and apply this small correction to the 
HS compilation data. 

The HS\markcite{HS90} $logN-logS$ data plotted in Figure 7 rely at 
their faintest end---overlapping our bright end---especially on the 
work of Koo and Kron and collaborators (e.g., Koo \& Kron\markcite{KK88} 
1988). Based on the spectroscopically confirmed 
QSOs in the Koo \& Kron sample in the magnitude range 20.0$<B<$22.6 at 
0$<z<$2.1, a surface density of about 130 deg$^{\rm -2}$ may be inferred; 
the number of such QSOs predicted in our survey area using the latter 
value is then 3.2, and thus in good agreement with the 
2.0$^{\rm +2.6}_{\rm -1.3}$ such ``bright" QSO candidates we actually find 
(although the numbers are very small in our sample for such ``bright"
QSOs). Similar agreement is found at our
``bright" end between the observed surface densities of our 
candidates and measures for confirmed $z\lesssim 2.1$  QSOs in the 
Canada-France Redshift Survey (Schade et al.\markcite{Sch96} 1996) 
to $B<23$. The slope of the HS\markcite{HS90} cumulative $logN-logS$ 
relation at the faint end is 0.35 for $\lesssim2.1$ QSOs. An extrapolation to 
$B\lesssim24.5$ of the HS\markcite{HS90} slope predicts for our 
survey area 14.1 QSOs; this prediction agrees reasonably well with our 
measure from Sample I candidates of  $10.4^{+4.4}_{-3.2}$ (Table 6).
Thus, as may be discerned visually from Figure~7, a simple 
extrapolation of the HS\markcite{HS90} $z<2.1$ QSO $logN-logS$ curve to 
$B\lesssim24.5$ appears to also match our observed surface density measure of 
candidates very well. 

A popular model of the QSO luminosity function---at least for faint, low- to 
moderate-$z$ QSOs---has been that presented by Boyle et al.\markcite{BSP88} 
(1988; hereafter BSP), Boyle\markcite{Boy91} (1991), and see also 
Marshall\markcite{Mar87} (1987). They 
parameterize the luminosity function as a double power law, with a shallow 
component at low luminosities and a steep component at high luminosities. In 
the case of pure luminosity evolution, the redshift dependence is often 
expressed as a power law $L(z) \propto (1+z)^{k_{\rm L}}$, with 
$k_L\approx3.15$. We carry out an integration to $z=2.1$ of the 
BSP\markcite{BSP88} luminosity function
using their best-fit model parameters, and for $q_{\rm 0}$=0.5  and 
$H_{\rm 0}$=50 (their model~B), to predict the 
$logN-logS$ curve for QSOs to $B<24.5$, and beyond; $B=24.5$ corresponds to 
$M_B=-20.9$ at $z=2.1$ for their model~B. The predictions of the 
BSP\markcite{BSP88} 
luminosity function to $B\lesssim$24.5 (solid curve in Figure 7) are in good 
agreement with the observed surface densities of our Sample II candidates.
As may also be seen in Figure~7 (dashed curve), similarly good 
agreement is found between our faint candidate surface densities and 
the predictions of a Gaussian form of pure luminosity evolution proposed by 
Pei\markcite{Pei95} (1995); the Pei luminosity function fits the available 
$logN-logS$ curve data for brighter confirmed QSOs
reasonably well over the entire observed redshift range, even beyond $z>3$. 
Lastly, we plot in Figure 7 (dotted curve) the predictions of an example of
a physically-based model luminosity function proposed by Yi\markcite{Yi96} 
(1996); the redshift dependence of the QSO luminosity in a low-efficiency 
advection-dominated regime can be approximated in the Yi model by a 
standard pure luminosity evolution form, and again there is good agreement 
between the model predictions (for a plausible QSO formation epoch near
$z\approx3.6$; see Beck-Winchatz\markcite{BW98} 1998 for details) and our 
ultrafaint $logN-logS$ data points in Figure 7.

In summary, as is evident from Figure 7, our surface density of 
morphologically plus multicolor-selected QSO {\it candidates} to 
$m_{606}<24$ (corresponding approximately to $B\lesssim24.5$) agrees 
extremely well with extrapolations to faint magnitudes from a variety of 
other brighter QSO surveys, as well as with the predictions of several pure
luminosity function models, both phenomenological and physically-based. 

Finally, we also plot in Figure 7 a new upper limit derived from
the Conti et al.\markcite{Con99} (1999) study of even fainter 
QSO/AGN candidates in the Hubble Deep Field (hereafter HDF). 
Conti et al.\markcite{Con99} (1999)  independently arrive at a combined 
multicolor and morphological selection approach, although their morphological 
criteria for resolved QSO/AGNs may not be quite so specifically targeted at 
QSOs/AGNs with point-source nuclei as that we describe in \S 3.2.
Conti et al. ultimately quote an upper limit to the 
QSO/AGN surface density at $m_{606}<27.0$, as their criteria also
include objects having colors consistent with normal stars.
Their Tables 4 and 6 list 14 objects they call attention to as potential 
low-$z$ QSOs or resolved (in WFPC2) AGN candidates which,
for the HDF area of 4.4 square arcminutes, corresponds to 
an upper limit of $<$11,500 deg$^{-2}$ QSOs/AGNs with $m_{606}<27.0$,
corresponding to about $B\lesssim27.5$. There is no inconsistency between the 
upper limit inferred from the Conti et al.\markcite{Con99} (1999) 
data, and a smooth extrapolation from the HS\markcite{HS90} $logN-logS$ curve 
nor with our candidate points corresponding to $B\lesssim24.5$. 
It appears possible that some of the low-$z$ candidates in the HDF are not 
QSOs/AGNs (see additional discussion in Conti et al.\markcite{Con99} 1999), 
or that a large 
population of extremely low luminosity QSOs/AGNs has emerged in the HDF. 

\section{Summary and Conclusions}

We have completed a combined multicolor and morphological survey for 
QSOs to $m_{606}<24.0$ ($B\lesssim$24.5) in 0.025 deg$^2$ of the 
Groth-Westphal Strip, using $F606W$ and $F814W$ WFPC2 images from HST and 
$u^*$-band images from the APO 3.5-m. Objects are selected as QSO/AGN 
candidates according to their UV-excess (likely $z\lesssim2.1$), and if they 
have either stellar morphology, or an approximately stellar-nucleus even 
if an underlying galaxy is evident, in HST images. 

We find 10 high-likelihood QSO/AGN candidates in our survey area which satisfy 
these criteria (Sample I). We also devise a second sample (Sample II) which 
excludes three objects from Sample I on the basis of their red 
$(m_{606}-m_{814})$ color,
to potentially further reduce residual contamination by compact galaxies. 
The cumulative surface density of QSO candidates for $19.3< m_{606} <24.0$, the 
approximate equivalent of $19.8 \lesssim B \lesssim 24.5$, is 
420$^{\rm +180}_{\rm -130}$ deg$^{-2}$ for Sample I, or
290$^{\rm +160}_{\rm -110}$ deg$^{-2}$ for Sample II. 
At our ``bright" end ($B\lesssim22.5$) our $logN-logS$ relation for 
candidates agrees well with the surface density of QSOs found by 
Koo \& Kron\markcite{KK88} (1988), Schade et al.\markcite{Sch96}
(1996), and others. At the faint end, our results are consistent with  
extrapolations to $B\lesssim24.5$ 
of the HS\markcite{HS90} $logN-logS$ curve, and the predictions of 
several recent pure luminosity evolution models (BSP\markcite{BSP88};
Pei\markcite{Pei95} 1995; Yi\markcite{Yi96} 1996). Hence, this pure 
luminosity evolution description, which has often been argued to provide a 
good representation of the evolution of QSOs to $B<$22.5, may be a good 
description of the evolution of even fainter (low- to modest-redshift) QSOs as 
well.  Note that such good agreement with extrapolations 
from brighter surveys and model luminosity function predictions need not have 
been the case: as discussed in \S 3.2, we would have found nearly an order of 
magnitude more UV-excess quasar candidates had we used only color information
and neglected the morphology information available in HST images.

While it is gratifying that the surface densities derived from our QSO
candidates are in good agreement with extrapolations from brighter
spectroscopically confirmed QSO samples and model predictions, follow-up 
of our candidates is of
course highly desirable and well within the capabilities of 10m-class 
telescopes instrumented for multi-object spectroscopy. Spectroscopic 
confirmations of our candidates as QSOs/AGNs would allow an initial exploration 
of the ultrafaint luminosity function (and any confirmations of very
faint UV-excess white dwarfs would also be intriguing). Of course, even 
negative spectroscopic results indicating that the sample candidates are not 
QSOs/AGNs are potentially of high interest; such results would even more
stringently constrain our already low initial estimates of the surface
densities of ultrafaint QSOs.

\section{Acknowledgments}
We thank the University of Washington 3.5-m TAC for generous 
allocations of observing time, the excellent APO support staff, and 
C.~Stubbs, A.~Diercks, P.~Doherty, and E.~Magnier for their development of 
and shared expertise on SPIcam. We thank E.~Deutsch for creating IDL code 
used for some of data reductions herein. 
The Medium-Deep Survey analysis was funded by the HST WFPC2 Team and STScI 
grants GO2684, 
GO6951, and GO7536 to Prof. Richard Griffiths and Dr. Kavan Ratnatunga at 
Carnegie Mellon University. Our work was supported at the University of 
Washington by NASA/HST grant GO-07976.01-96A, and is based on 
observations with the NASA/ESA Hubble Space Telescope, obtained  at the Space 
Telescope Science Institute, which is operated by the Association of 
Universities for Research in Astronomy, Inc., under NASA contract NAS5-26555.


\begin{table}
\begin{center}
Table 1. UV-Excess (\& Misc. High-$z$) QSO Candidates with Stellar-PSFs in WFPC2
\begin{tabular}{rccccccc}
\tableline
\tableline 
Cat. & RA        & Dec            & $m_{606}$ & $u^*-m_{606}$ & $m_{606}-m_{814}$ & Stellarity & Comments\\
 ID  & (14:)     & (+52:)         & (Aper)    & (Aper)        & (Aper)  &   &        \\
\tableline 
  98 & 15:30.49 & 04:39.9 & 19.88 &    0.59 & 0.36 & 0.89,0.90 & UV-excess \\ 
 971 & 16:13.41 & 11:36.3 & 20.60 &    0.41 & 0.37 & 0.98,1.00 & UV-excess \\ 
 815 & 16:29.00 & 15:33.0 & 23.61 &   -0.14 & 0.46 & 0.98,1.00 & UV-excess \\ 
 245 & 17:28.35 & 28:13.3 & 23.91 &   -0.09 & 0.66 & 0.99,0.92 & UV-excess \\ 
     &                    &       &         &      &           &     \\ %
1082 & 15:42.65 & 09:27.6 & 21.21 &    2.31 & 0.22 & 1.00,1.00 & high-$z$? \\ 
1702 & 16:53.98 & 20:43.4 & 23.57 & $>$2.05 & 0.34 & 1.00,0.98 & high-$z$? \\ 
\tableline
\end{tabular}
\tablecomments{Column 1 - our catalog ID number; Columns 2 \& 3 - J2000 
coordinates; Column 7 - SExtractor stellarity in F606W,F814W.}
\end{center}
\end{table}

\begin{table}
\begin{center}
Table 2. UV-Excess QSO Candidates, Approx. Stellar-Nuclei in WFPC2
\begin{tabular}{rcccccc}
\tableline 
\tableline 
Cat. & RA    & Dec    & $m_{606}$ & $u^*-m_{606}$ & $m_{606}-m_{814}$ & $I_{RAD}$, \\
 ID  & (14:) & (+52:) & (MDS)     & (MDS)         & (MDS)             & $R_{MAJ}\over{R_{MIN}}$       \\
\tableline 
  72 & 15:39.39 & 04:44.5 & 23.90 & 0.50 & 0.47 & 0.39,1.32 \\ 
1148 & 15:51.93 & 08:30.5 & 23.78 & 0.32 & 0.79 & 0.27,1.11 \\ 
1025 & 16:09.02 & 10:50.9 & 23.32 & 0.44 & 0.43 & 0.41,1.19 \\ 
 507 & 16:53.28 & 21:03.8 & 22.69 & 0.26 & 1.23 & 0.39,1.09 \\ 
 608 & 16:55.00 & 18:38.5 & 23.78 & 0.15 & 0.53 & 0.49,1.09 \\ 
 315 & 17:14.41 & 25:46.4 & 23.18 & 0.18 & 0.30 & 0.40,1.24 \\ 
 390 & 17:19.32 & 23:13.2 & 22.79 & 0.70 & 0.48 & 0.41,1.23 \\ 
\tableline 
\end{tabular}
\tablecomments{Column 1 - our catalog ID number; 
Columns 2 \& 3 - J2000 coordinates; Column 7 - WFPC2 radial profile integral,
major-to-minor axis ratio.}
\end{center}
\end{table}

\begin{table}
\begin{center}
Table 3. Miscellaneous Compact UV-Excess Objects (Too Extended or Too Faint)
\begin{tabular}{rccccccc}
\tableline 
\tableline 
Cat. &  RA   &  Dec   & $m_{606}$ & $u^{*}-m_{606}$ & $m_{606}-m_{814}$ & $I_{RAD}$, & Comments \\
 ID  & (14:) & (+52:) & (MDS)     & (MDS)           & (MDS)             & $R_{MAJ}\over{R_{MIN}}$ & \\
\tableline 
1233 & 15:14.74 & 01:37.6 & 23.62 &  0.38 & 0.41 & 0.56,1.27 & slight extension \\ 
1176 & 15:15.88 & 03:27.0 & 24.06 & -0.14 & 0.67 & 0.45,1.24 & $m_{606}>24$\\ 
 802 & 16:22.65 & 15:07.5 & 23.94 &  0.30 & 0.36 & 0.56,1.12 & slight extension \\ 
 757 & 16:36.54 & 16:49.9 & 23.60 &  0.11 & 0.77 & 0.60,1.01 & slight extension \\ 
 505 & 16:51.34 & 20:46.1 & 21.37 &  0.21 & 0.41 & 0.57,1.07 & face-on host\\ 
 359 & 17:10.90 & 23:46.6 & 23.88 &  0.63 & 0.54 & 0.77,1.94 & tidal tail?\\ 
 163 & 17:54.64 & 30:57.0 & 23.20 & -0.07 & 0.76 & 0.67,1.42 & edge-on host?\\ 
\tableline 
\end{tabular}
\tablecomments{Column 1 - our catalog ID number; Columns 2 \& 3 - J2000 
coordinates; Column 7 - WFPC2 radial profile integral, major-to-minor axis 
ratio; Column 8 - comments (mainly on visual appearance in WFPC2).}
\end{center}
\end{table}

\begin{table}
\begin{center}
Table 4. Summary Information for QSO Candidates in Combined UV-Excess Sample I
\begin{tabular}{rrccccccc}
\tableline 
\tableline 
Obj. & Cat. &  RA   &  Dec   & $m_{606}$  & Stellarity & MDS   & $I_{RAD}$, & Comments \\
 No. &  ID  & (14:) & (+52:) & (Aper)     &            & Class & $R_{MAX}\over{R_{MIN}}$ & \\
\tableline 
 1 &   98 & 15:30.49 & 04:39.9 & 19.88 & 0.89,0.90 & 0,3 & 0.21,1.12 & obvious host \\ 
 2 & 1148 & 15:51.93 & 08:30.5 & 23.83 & 0.61,0.34 & 0,1 & 0.27,1.11 &  \\ 
 3 & 1025 & 16:09.02 & 10:50.9 & 23.31 & 0.77,0.37 & 0,2 & 0.41,1.19 & obvious host \\ 
 4 &  971 & 16:13.41 & 11:36.3 & 20.60 & 0.98,1.00 & 0,0 & 0.01,1.23 &  \\  
 5 &  815 & 16:29.00 & 15:33.0 & 23.61 & 0.98,1.00 & 0,0 & $-$    & nearby spiral \\ 
 6 &  507 & 16:53.28 & 21:03.8 & 22.93 & 1.00,0.04 & 2,2 & 0.39,1.09 & fuzz in F814W \\  
 7 &  608 & 16:55.00 & 18:38.5 & 23.87 & 0.27,0.06 & 1,1 & 0.49,1.09 &  \\ 
 8 &  315 & 17:14.41 & 25:46.4 & 23.29 & 0.03,0.04 & 1,1 & 0.40,1.24 &  \\ 
 9 &  390 & 17:19.32 & 23:13.2 & 22.79 & 0.56,0.44 & 0,2 & 0.41,1.23 & fuzz;close star? \\ 
10 &  245 & 17:28.35 & 28:13.3 & 23.91 & 0.99,0.92 & 2,2 & 0.49,1.23 & fuzz? in F814W \\     
\tableline 
\end{tabular}
\tablecomments{Column 1 - QSO candidate no. in Beck-Winchatz (1998);
Column 2 - our catalog ID number; Columns 3 \& 4 - J2000 
coordinates; Column 6 - SExtractor stellarity in F606W,F814W; 
Column 7 - MDS morphological class; Column 8 - WFPC2 radial profile integral, 
major-to-minor axis ratio; Column 9 - comments on visual appearance in WFPC2.}
\end{center}
\end{table}

\begin{table}
\begin{center}
Table 5. Object Detection Probabilities for SPIcam Images
\begin{tabular}{ccccc}
\tableline 
\tableline 
$u^*$ &  2nd Deepest &  2nd Shallowest &  Typical Value   \\  
      &  Field       &  Field          & \\
\tableline 
23.0-23.5 & 0.97 & 0.94 & 0.97 \\
23.5-24.0 & 0.96 & 0.92 & 0.96 \\
24.0-24.5 & 0.91 & 0.78 & 0.83 \\
24.5-24.7 & 0.61 & 0.41 & 0.49 \\
\tableline 
\end{tabular}
\end{center}
\end{table}

\begin{table}
\begin{center}
Table 6. Cumulative Surface Densities of QSO Candidates
\begin{tabular}{cccccc}
\tableline 
\tableline 
$m_{606}$ & $B$ &  n     & N (deg$^{-2}$) &  n       & N (deg$^{-2}$) \\  
          &     & Sample I & Sample I      & Sample II  & Sample II  \\  
\tableline 
23.0 & 23.5 & 5.0$^{\rm +3.4}_{\rm -2.2}$   & 210$^{\rm +140}_{\rm -90}$ & 
3.0$^{\rm +3.0}_{\rm -1.6}$ & 120$^{\rm +120}_{\rm -70}$ \\
23.5 & 24.0 & 7.1$^{\rm +3.8}_{\rm -2.6}$  & 290$^{\rm +160}_{\rm -110}$ & 
5.1$^{\rm +3.4}_{\rm -2.2}$ & 210$^{\rm +140}_{\rm -90}$ \\
24.0 & 24.5 & 10.4$^{\rm +4.4}_{\rm -3.2}$  & 420$^{\rm +180}_{\rm -130}$ & 
7.2$^{\rm +3.8}_{\rm -2.6}$ & 290$^{\rm +160}_{\rm -110}$ \\
\tableline 
\end{tabular}
\tablecomments{Columns 3 \& 5 - cumulative number of QSO candidates in survey 
region, corrected for detection incompleteness; Columns 4 \& 6 - cumulative 
surface density of QSO candidates, corrected for detection incompleteness.}
\end{center}
\end{table}

\clearpage

\begin{figure}
\plotone{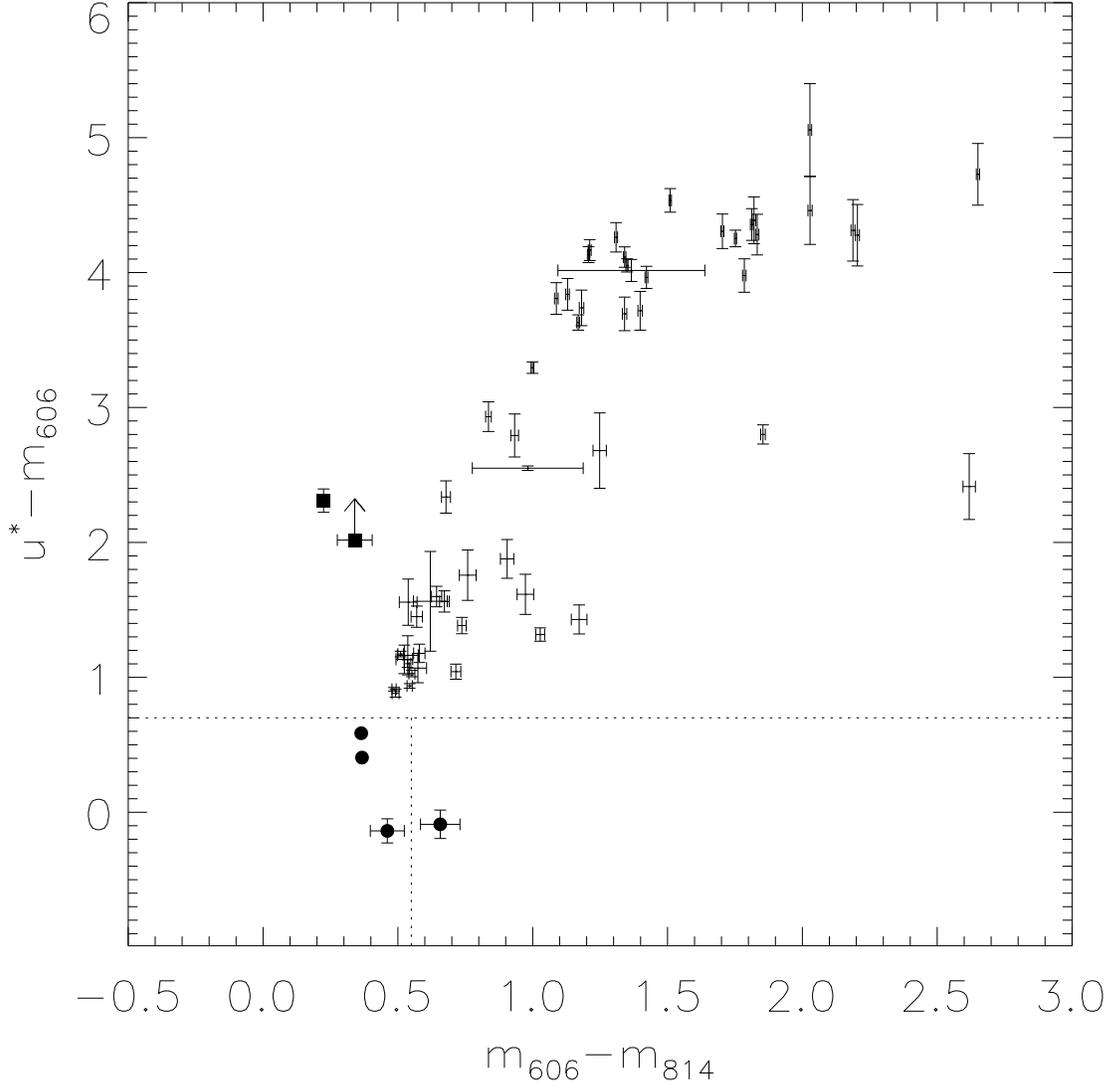}
\figcaption{Two-color diagram of objects (1$\sigma$ photometric errors) in our 
survey to an aperture magnitude limit of $m_{606}<24$, that were classified as 
having stellar morphology in WFPC2 by the SExtractor neural network. For 
added clarity, we do not plot $u^*$ non-detections (with 
one exception). Filled circles denote the four UV-excess quasar candidates 
(see also Table 1) with stellar-PSF morphology in WFPC2
(for two of these candidates, the photometric errors are merely smaller than 
the plot symbol). The dotted horizontal and vertical lines indicate, 
respectively, our UV-excess criterion and an additional possible 
$(m_{606}-m_{814})$ color-cut to further exclude compact emission line 
galaxies. The filled squares denote two miscellaneous higher redshift QSO 
candidates (see Table 1).}
\end{figure}

\newpage

\begin{figure}
\plotone{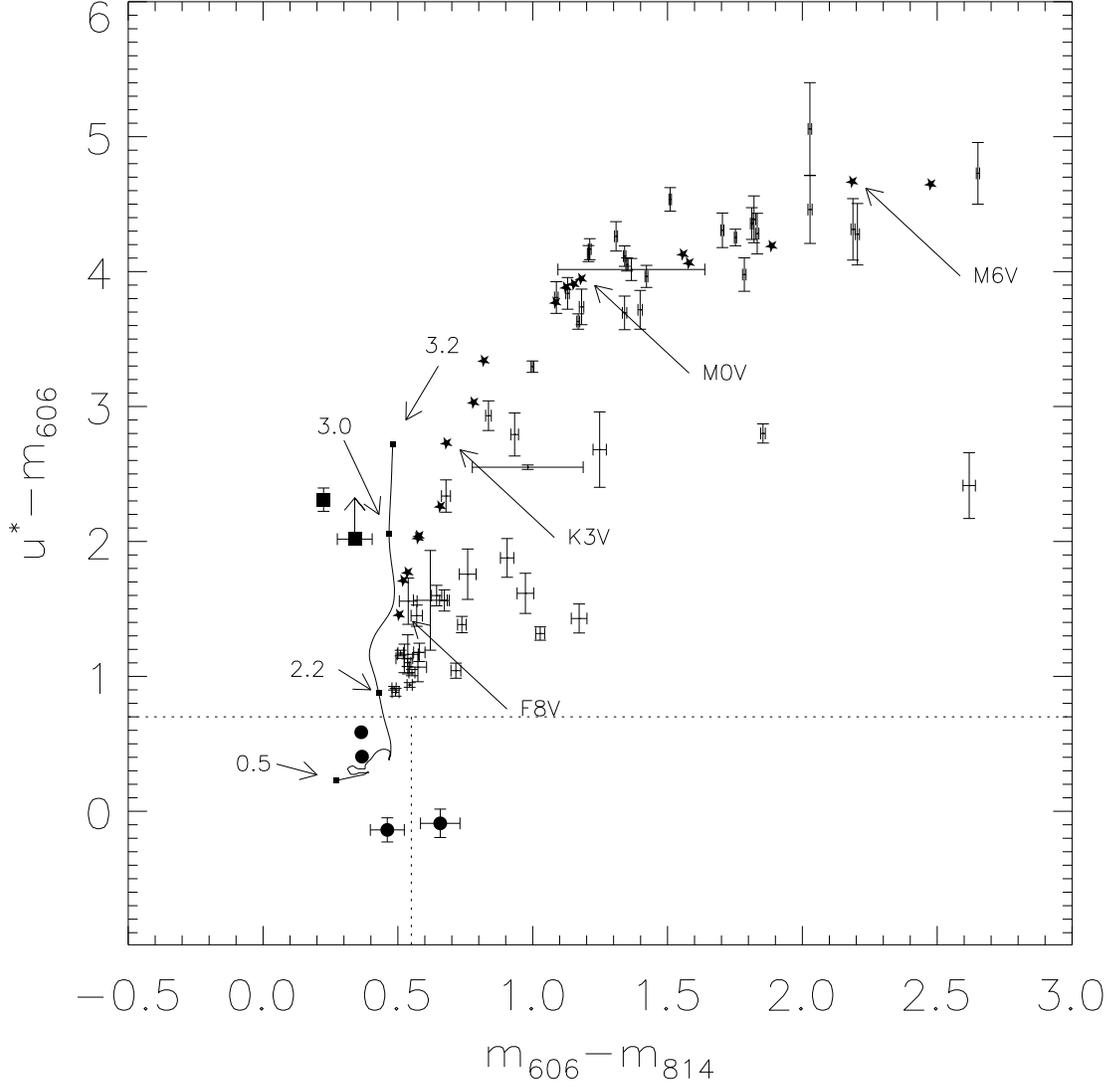}
\figcaption{Two-color diagram of stellar-PSF (in WFPC2) objects in our survey 
to an aperture magnitude limit of $m_{606}<24$. Data and symbols are
identical to those shown in Figure 1, except that we also overplot
for rough comparison the locus of colors expected of a typical quasar at 
various redshifts (solid curve), and the approximate colors predicted for
main sequence stars (5-point star symbols) later than spectral type F8.}
\end{figure}

\newpage

\begin{figure}
\plotone{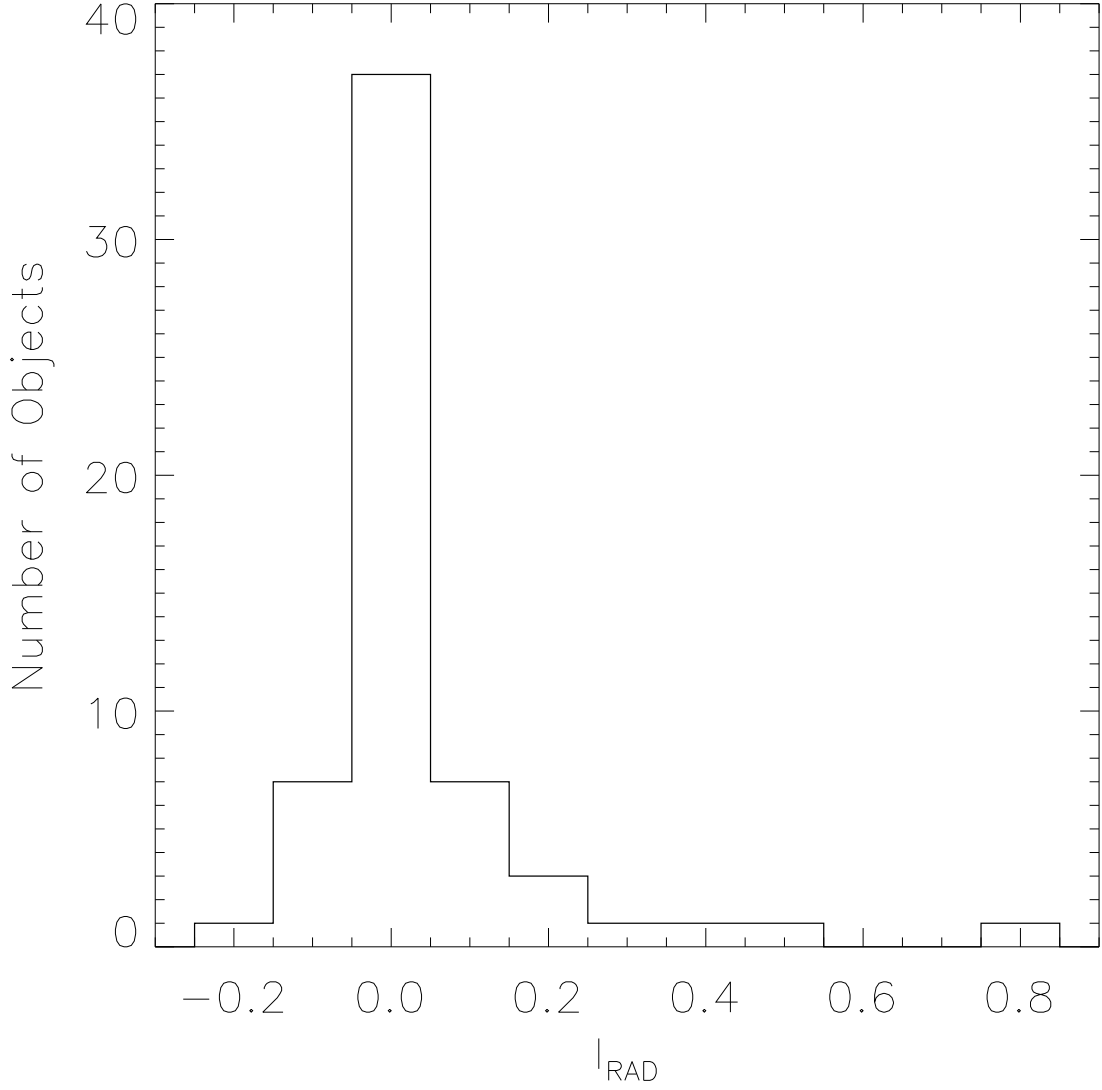}
\figcaption{Histogram of radial profile integrals, $I_{RAD}$, for WFPC2 objects
having stellar-PSFs according to SExtractor (distribution of
$I_{RAD}$ for objects plotted in Figure 1). As nearly all such stellar-PSF in 
WFPC2 objects have $I_{RAD}<0.5$, we also invoke this criterion as one 
portion of a dual quantitative test to assess whether UV-excess galaxies 
resolved in WFPC2 images, nonetheless have an approximately stellar-PSF
nucleus, and hence may be high likelihood QSO/AGN candidates.}
\end{figure}

\begin{figure}
\plotone{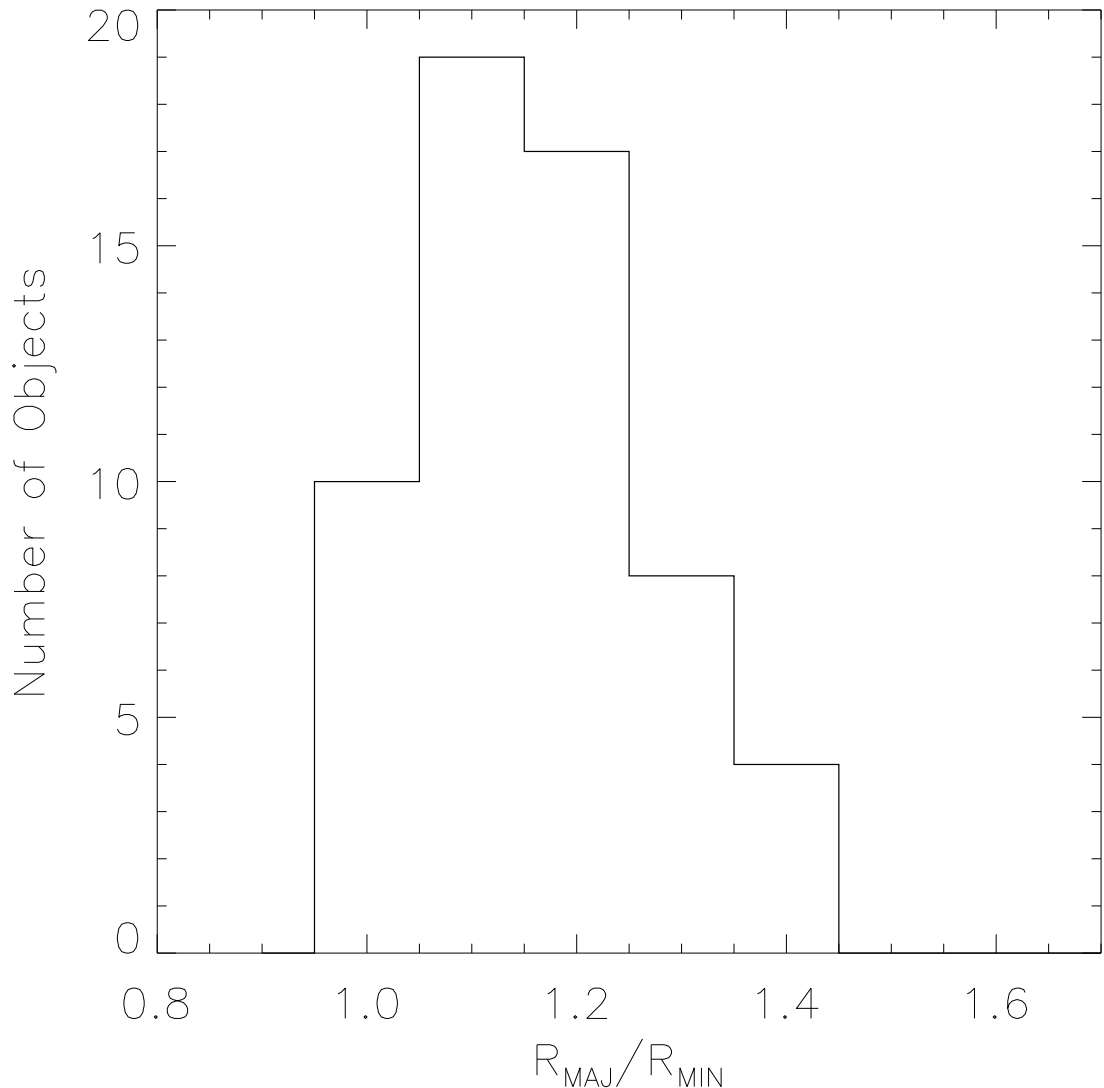}
\figcaption{Histogram of the major to minor axis ratios ($R_{MAJ}/R_{MIN}$) of 
two-dimensional elliptical Gaussian fits to WFPC2 objects
having stellar-PSFs according to SExtractor (distribution of
$R_{MAJ}/R_{MIN}$ for objects plotted in Figure 1). 
As all such stellar-PSF in WFPC2
objects have $R_{MAJ}/R_{MIN}<1.5$, we also invoke this
criterion as the second portion of a dual quantitative test to 
assess whether UV-excess galaxies resolved in WFPC2 images,
nonetheless have an approximately stellar-PSF nucleus, and hence may be
high likelihood QSO/AGN candidates.}
\end{figure}

\begin{figure}
\plotone{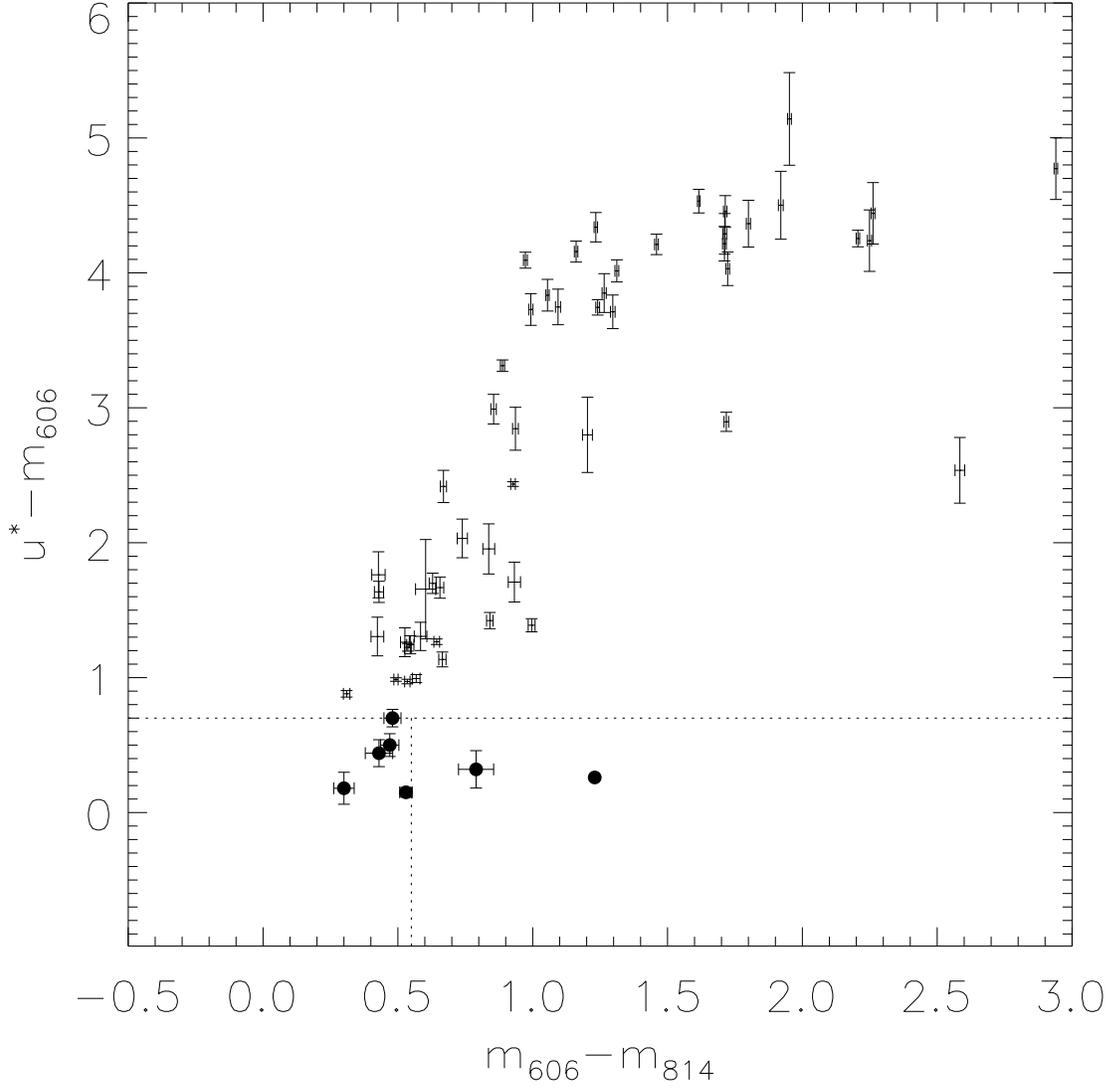}
\figcaption{Two-color diagram of objects (1$\sigma$ photometric errors) in our 
survey to an MDS magnitude limit of $m_{606}<24$. Error bars denote
colors of objects classified as having stellar-PSF morphology in WFPC2
by the MDS algorithm (but otherwise this figure is analogous to Figure 1).
The filled circles overplot the colors of additional UV-excess 
QSO/AGN candidates (see Table 2) that have stellar-nuclei+galaxy morphology. 
(QSO candidates already plotted in Figures 1--2 and listed in Table 1
are {\it not} replotted here).}
\end{figure}

\begin{figure}
\plotone{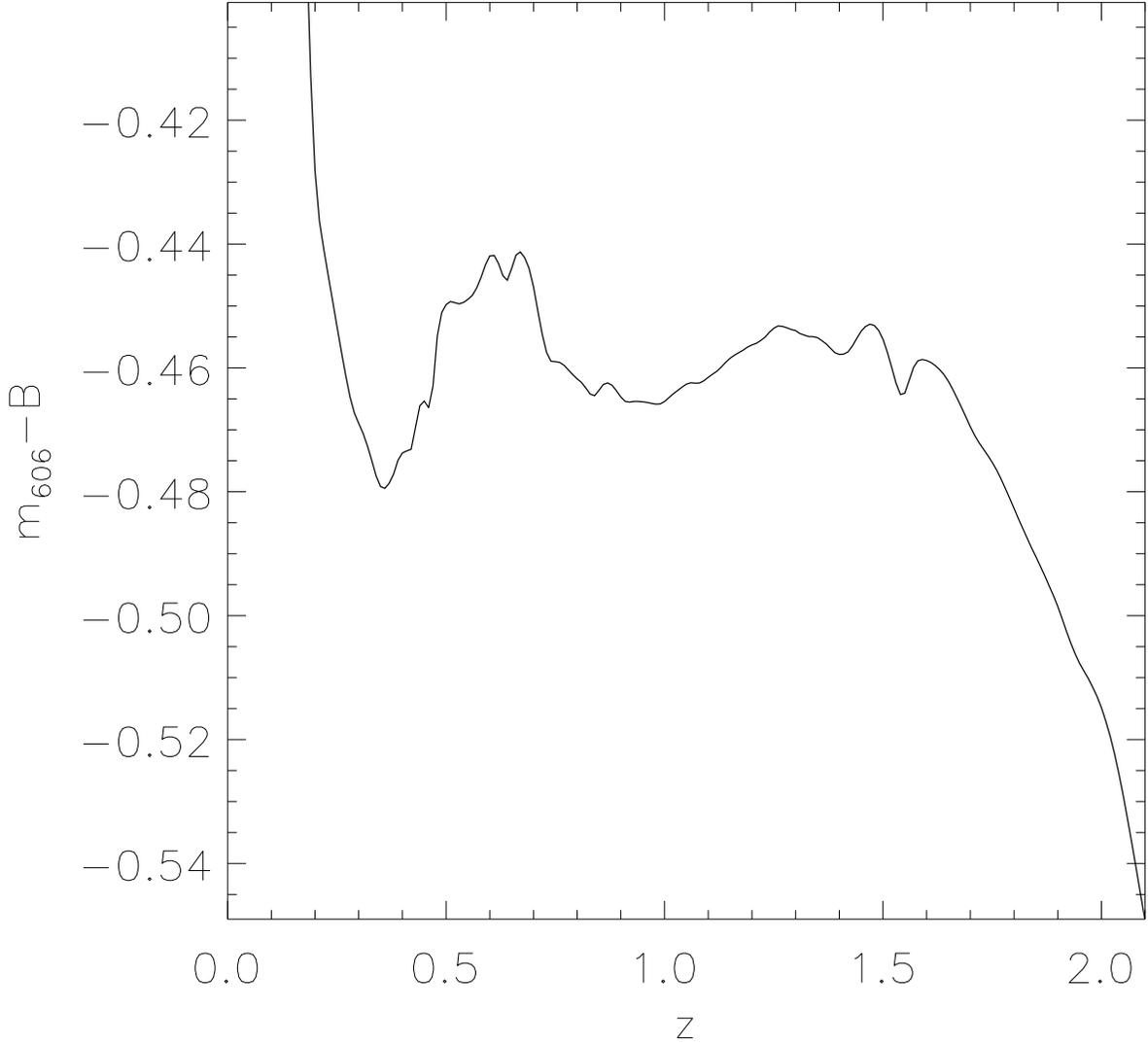}
\figcaption{A rough conversion, as a function of redshift, between
the $F606W$ magnitude (including the contribution of emission lines) of a 
typical QSO and the Johnson $B$-band continuum magnitude.
This was estimated by convolving the redshifted composite LBQS 
spectrum, as well as a cubic spline fit to its continuum, with the wavelength 
dependent sensitivity curves of the HST/WFPC2/$F606W$ system and the $B$-band. 
The LBQS spectrum is defined for $\lambda\leq$6000 \AA\ 
and does not completely overlap the $F606W$ bandpass for z$\lesssim$0.2; this 
causes the sharp increase in $(m_{606}-B)$ at low redshifts. For very
approximate comparison with the results of other quasar surveys defined in 
terms of $B$-band magnitudes, we adopt $B\approx m_{606}+0.5$ as typical 
for $z<2.1$ QSOs.}
\end{figure}

\begin{figure}
\plotone{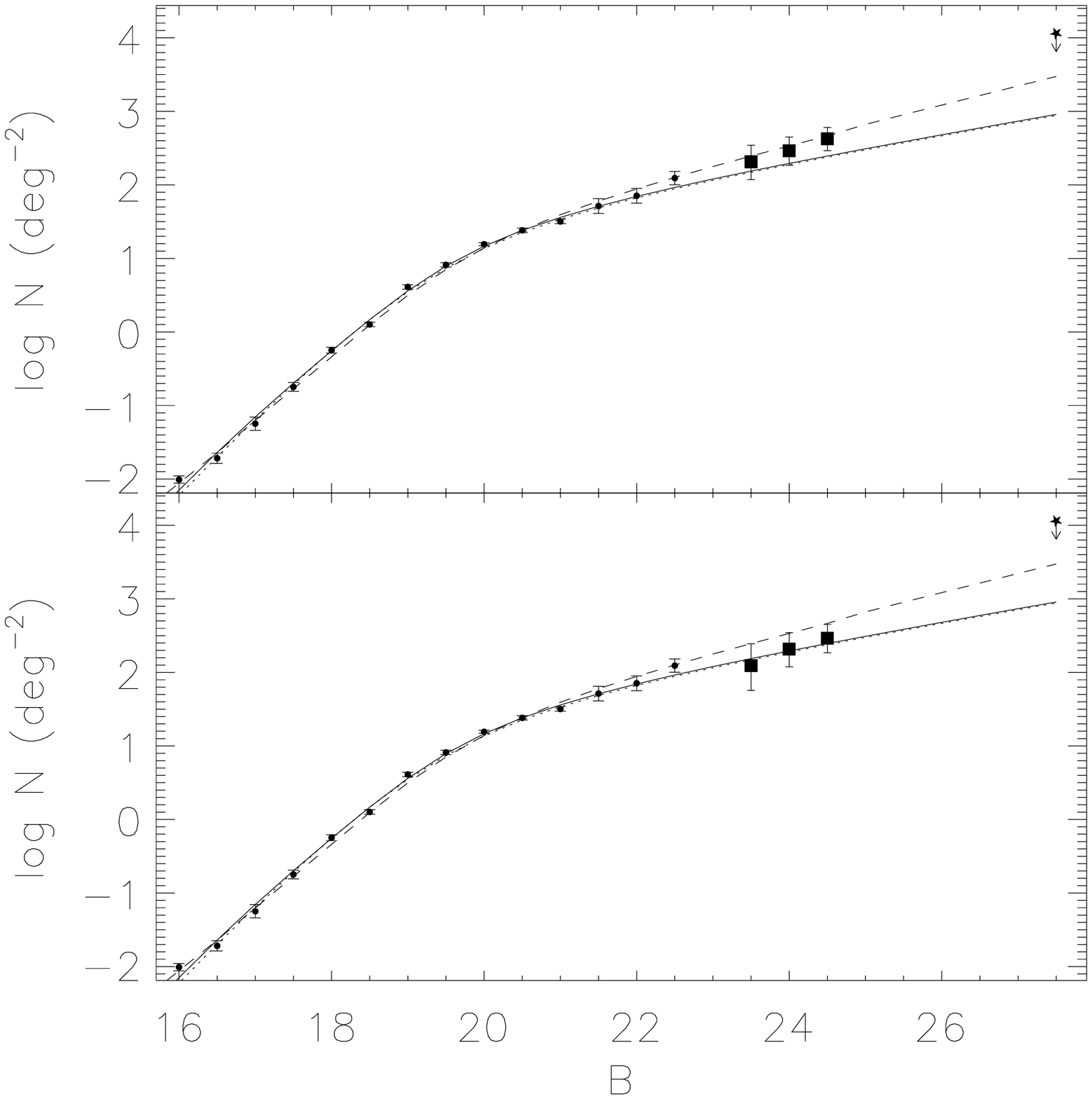}
\figcaption{Comparison of cumulative surface densities of QSO/AGN candidates
(squares) from Sample I (upper panel) and Sample II (lower panel) with the 
predictions of three pure luminosity evolution models for $z<2.1$. Solid
curves
show the predictions of a double power-law luminosity function combined with 
the power-law pure luminosity evolution given by BSP. Dashed curves show the 
predictions of a Gaussian pure luminosity evolution of Pei (1995), which
fits even higher-$z$ QSO data as well. Dotted curves show the predictions of 
a physically based model by Yi (1996). Also plotted (small circles) are the 
surface densities of confirmed brighter QSOs derived by HS, as well as an 
upper limit (arrow at $B\approx27.5$) on extremely faint low-$z$ QSOs from 
the HDF (Conti et al. 1999). Excellent agreement is found between the surface
density of QSO candidates in our samples to $B\lesssim24.5$,
predictions of pure luminosity evolution models, and simple extrapolations 
from brighter surveys.}
\end{figure}


\newpage
\begin{references}

\reference{Bah97} Bahcall, J. N., Kirhakos, S., Saxe, D. H., \& 
Schneider, D. P. 1997, ApJ, 479, 642

\reference{BW98} Beck-Winchatz, B. 1998, Ph.D. Thesis, University of
Washington

\reference{BA96} Bertin, E., \& Arnouts, S. 1996, A\&AS, 117, 393

\reference{Blan86} Blandford, R. 1986, in Quasars: Proc. of the 119th
Symp. of the IAU, ed. G. Swarup \& V. K. Kapahi (Dordrecht: Reidel), 359

\reference{BSP88} Boyle, B. J., Shanks, T., \& Peterson, B. A. 1988, MNRAS, 
235, 935 (BSP)

\reference{Boy91} Boyle, B. J. 1991, in The Space Distribution of Quasars,
ASP Conference Series Vol. 21, ed. D. Crampton (San Francisco: ASP), 389

\reference{Bru99} Brunner, R. J., Connolly, A. J., \& Szalay, A. S. 1999, 
ApJ, in press

\reference{Cav89} Cavaliere \& Padovani 1989, ApJ, 340, L5

\reference{Con99} Conti, A., Kennefick, J. D., Martini, P., \& Osmer, P. S. 
1999, AJ, in press 

\reference{Fle86} Fleming, T. A., Liebert, J., \& Green, R. F. 1986, 
ApJ, 308, 176

\reference{Fran91} Francis, P. J., Hewett, P. C., Foltz, C. B., Chaffee,
F. H., Weymann, R. J., \& Morris, S. L. 1991, ApJ, 373, 465

\reference{Fuk96} Fukugita, M., Ichikawa, T., Gunn, J. E., Doi, M.,
Shimasaku, K., \& Schneider, D. P. 1996, AJ, 111, 1748

\reference{Gou98} Gould, A., Flynn, C., \& Bahcall, J. N. 1998, 
ApJ, 503, 798

\reference{Gra98} Graham, M. J., Clowes, R. G., \& Campusano, L. E. 1998,
ApJ, in press 

\reference{Gri94} Griffiths, R. E. et al. 1994, ApJ, 437, 67

\reference{Gro94} Groth, E. J., Kristian, J. A., Lynds, R., O'Neil, E. J.,
Balsano, R., Rhodes, J., \& the WFPC-1 IDT 1994, BAAS, 26, 1403

\reference{GS83} Gunn, J. E., \& Stryker, L. 1983, ApJS, 52, 121

\reference{Hai99} Haiman, Z., \& Menou, K. 1999, ApJ, submitted


\reference{Hall96} Hall, P. B. , Osmer, P. S., Green, R. F., Porter, A. C.,
\& Warren, S. J. 1996, ApJ, 462, 614

\reference{Han98} Hansen, B., 1998, Nature, 394, 860

\reference{HS90} Hartwick, F. D. A., \& Schade, D. 1990, ARAA, 28, 437 (HS)

\reference{Hol95} Holtzman, J. A., Burrows, C. J., Casertano, S., Hester,
J. J., Trauger, J. T., Watson, A. M., \& Worthey, G. 1995, PASP, 107, 1065

\reference{Hoo97} Hooper, E. J., Impey, C. D., \& Foltz, C. 1997, ApJ, 480, L95

\reference{Hut94} Hutchings, J. B., Holtzman, J., Sparks, W. B., Morris, S. C.,
Hanisch, R. J. 1994, ApJ, 429, L1

\reference{JV99} Jarvis, R. M. \& MacAlpine, G. M. 1999, AJ, in press

\reference{Ken97} Kennefick, J. D., Osmer, P. S., Hall, P. B., \& Green, R. F.
1997, AJ, 114, 2269

\reference{KKC86} Koo, D. C., Kron, R. G. \& Cudworth, K. M. 1986, PASP, 98, 285

\reference{KK88} Koo, D. C., \& Kron, R. G. 1988, ApJ, 325, 92

\reference{Koo96} Koo, D. C., et al. 1996, ApJ, 469, 535

\reference{Lie88} Liebert, J., Dahn, C. C., \& Monet, D. G. 1988, 
ApJ, 332, 891

\reference{Mal98} Malkan, M. A., Gorjian, V., \& Tam, R. 1998, ApJS, 117, 25

\reference{Mar87} Marshall, H. L. 1987, AJ, 94, 628

\reference{Osm98} Osmer, P. S., Kennefick, J. D., Hall, P. B., \& 
Green, R. F. 1998, ApJS, in press 

\reference{Pei95} Pei, Y. C. 1995, ApJ, 438, 623

\reference{Rat99} Ratnatunga, K. U., Griffiths, R. E., \& Ostrander E. J.
1999, in preparation

\reference{Reg51} Regener, V. H. 1951, Phys. Rev., 84. L161

\reference{Roc96} Roche, N., Ratnatunga, K., Griffiths, R. E., Im, M., \&
Neuschaefer, L. 1996, MNRAS, 282, 1247

\reference{Sar96} Sarajedini, V. L., Green, R. F., Griffiths, R. E., \&
Ratnatunga K. U. 1996, ApJ, 471, L15

\reference{Sar99a} Sarajedini, V. L., Green, R. F., Griffiths, R. E., \&
Ratnatunga K. U. 1999a, ApJS, in press

\reference{Sar99b} Sarajedini, V. L., Green, R. F., Griffiths, R. E., \&
Ratnatunga K. U. 1999b, ApJ, in press

\reference{Sch96} Schade, D., Crampton, D., Hammer, F., Le Fevre, O., 
\& Lilly, S. J. 1996, MNRAS, 278, 95

\reference{Sch83} Schmidt, M. \& Green, R. F. 1983, ApJ, 269, 352

\reference{Ver83} Veron, P. 1983, in Quasars and Gravitational Lenses, Proc. 
of the 24th Liege International Astrophysical Colloquium, 
ed. J. P. Swings (Li\'{e}ge: Univ. de Li\'{e}ge), 210

\reference{War87} Warren, S. J.,  Hewett, P. C.,  Irwin, M. J.,  McMahon, R. G.,
Bridgeland, M. T. , Bunclark, P. S. , \& Kibblewhite, E. J. 1987, 
Nature, 325, 131

\reference{Whit95} Whitmore, B. 1995, in Calibrating Hubble Space Telescope: 
Post Servicing Mission, ed. A. Koratkar \& C. Leitherer (Baltimore: STScI),
269

\reference{Yi96} Yi, I. 1996, ApJ, 473, 645



\end{references}
\end{document}